\input harvmac

\input amssym

\def\unit{\relax{\rm 1\kern-.26em I}}
\def\nada{\relax{\rm 0\kern-.30em l}}
\def\tilde{\widetilde}


\mathchardef\square"2299

\noblackbox
\def\IL{\relax{\rm I\kern-.18em L}}
\def\IH{\relax{\rm I\kern-.18em H}}
\def\IR{\relax{\rm I\kern-.18em R}}
\def\IC{\relax\hbox{$\inbar\kern-.3em{\rm C}$}}
\def\IZ{\relax\ifmmode\mathchoice
{\hbox{\cmss Z\kern-.4em Z}}{\hbox{\cmss Z\kern-.4em Z}} {\lower.9pt\hbox{\cmsss Z\kern-.4em Z}}
{\lower1.2pt\hbox{\cmsss Z\kern-.4em Z}}\else{\cmss Z\kern-.4em Z}\fi}

\def\CR {{\cal R}}

\def\partialslash{\not{\hbox{\kern-2pt $\partial$}}}

\def\CL {{\cal L}}

\def\CO {{\cal O}}

\def\CA{{\cal A}}


\def\CO {{\cal O}}

\font\manual=manfnt \def\dbend{\lower3.5pt\hbox{\manual\char127}}

\def\IZ{\relax\ifmmode\mathchoice
{\hbox{\cmss Z\kern-.4em Z}}{\hbox{\cmss Z\kern-.4em Z}} {\lower.9pt\hbox{\cmsss Z\kern-.4em Z}}
{\lower1.2pt\hbox{\cmsss Z\kern-.4em Z}}\else{\cmss Z\kern-.4em Z}\fi}

\def\bar{\overline}

\def\rt2{\sqrt{2}}
\def\irt2{{1\over\sqrt{2}}}

\def\slashchar#1{\setbox0=\hbox{$#1$}           
   \dimen0=\wd0                                 
   \setbox1=\hbox{/} \dimen1=\wd1               
   \ifdim\dimen0>\dimen1                        
      \rlap{\hbox to \dimen0{\hfil/\hfil}}      
      #1                                        
   \else                                        
      \rlap{\hbox to \dimen1{\hfil$#1$\hfil}}   
      /                                         
   \fi}

\def\foursqr#1#2{{\vcenter{\vbox{
    \hrule height.#2pt
    \hbox{\vrule width.#2pt height#1pt \kern#1pt
    \vrule width.#2pt}
    \hrule height.#2pt
    \hrule height.#2pt
    \hbox{\vrule width.#2pt height#1pt \kern#1pt
    \vrule width.#2pt}
    \hrule height.#2pt
        \hrule height.#2pt
    \hbox{\vrule width.#2pt height#1pt \kern#1pt
    \vrule width.#2pt}
    \hrule height.#2pt
        \hrule height.#2pt
    \hbox{\vrule width.#2pt height#1pt \kern#1pt
    \vrule width.#2pt}
    \hrule height.#2pt}}}}
\def\psqr#1#2{{\vcenter{\vbox{\hrule height.#2pt
    \hbox{\vrule width.#2pt height#1pt \kern#1pt
    \vrule width.#2pt}
    \hrule height.#2pt \hrule height.#2pt
    \hbox{\vrule width.#2pt height#1pt \kern#1pt
    \vrule width.#2pt}
    \hrule height.#2pt}}}}
\def\sqr#1#2{{\vcenter{\vbox{\hrule height.#2pt
    \hbox{\vrule width.#2pt height#1pt \kern#1pt
    \vrule width.#2pt}
    \hrule height.#2pt}}}}
\def\square{\mathchoice\sqr65\sqr65\sqr{2.1}3\sqr{1.5}3}

\def\dal{\raise1pt\hbox{$\sqr{6}{6}$}\,}
\def\figin{\epsfcheck\figin}\def\figins{\epsfcheck\figins}
\def\epsfcheck{\ifx\epsfbox\UnDeFiNeD
\message{(NO epsf.tex, FIGURES WILL BE IGNORED)}
\gdef\figin##1{\vskip2in}\gdef\figins##1{\hskip.5in}
\else\message{(FIGURES WILL BE INCLUDED)}%
\gdef\figin##1{##1}\gdef\figins##1{##1}\fi}
\def\DefWarn#1{}
\def\figinsert{\goodbreak\midinsert}
\def\ifig#1#2#3{\DefWarn#1\xdef#1{fig.~\the\figno}
\writedef{#1\leftbracket fig.\noexpand~\the\figno}%
\figinsert\figin{\centerline{#3}}\medskip\centerline{\vbox{\baselineskip12pt \advance\hsize by
-1truein\noindent\footnotefont{\bf Fig.~\the\figno:\ } \it#2}}
\bigskip\endinsert\global\advance\figno by1}


\lref\nelsonseiberg{
  A.~E.~Nelson and N.~Seiberg,
  ``R symmetry breaking versus supersymmetry breaking,''
  Nucl.\ Phys.\  B {\bf 416}, 46 (1994)
  [arXiv:hep-ph/9309299].
}

\lref\nonlinear{
  Z.~Komargodski and N.~Seiberg,
  ``From Linear SUSY to Constrained Superfields,''
  arXiv:0907.2441 [hep-th].
}.

\lref\nima{
  A.~Adams, N.~Arkani-Hamed, S.~Dubovsky, A.~Nicolis and R.~Rattazzi,
  ``Causality, analyticity and an IR obstruction to UV completion,''
  JHEP {\bf 0610}, 014 (2006)
  [arXiv:hep-th/0602178].}

\lref\IntriligatorPY{
  K.~A.~Intriligator, N.~Seiberg and D.~Shih,
  ``Supersymmetry Breaking, R-Symmetry Breaking and Metastable Vacua,''
  JHEP {\bf 0707}, 017 (2007)
  [arXiv:hep-th/0703281].
}

\lref\RayWK{
  S.~Ray,
  ``Some properties of meta-stable supersymmetry-breaking vacua in Wess-Zumino
  models,''
  Phys.\ Lett.\  B {\bf 642}, 137 (2006)
  [arXiv:hep-th/0607172].
}

\lref\KomargodskiJF{
  Z.~Komargodski and D.~Shih,
  ``Notes on SUSY and R-Symmetry Breaking in Wess-Zumino Models,''
  JHEP {\bf 0904}, 093 (2009)
  [arXiv:0902.0030 [hep-th]].
}

\lref\AffleckVC{
  I.~Affleck, M.~Dine and N.~Seiberg,
 ``Dynamical Supersymmetry Breaking In Chiral Theories,''
  Phys.\ Lett.\  B {\bf 137}, 187 (1984).
}

\lref\AffleckMF{
  I.~Affleck, M.~Dine and N.~Seiberg,
  ``Exponential Hierarchy From Dynamical Supersymmetry Breaking,''
  Phys.\ Lett.\  B {\bf 140}, 59 (1984).
}

\lref\ggmimpl{
  L.~M.~Carpenter, M.~Dine, G.~Festuccia and J.~D.~Mason,
  ``Implementing General Gauge Mediation,''
  Phys.\ Rev.\  D {\bf 79}, 035002 (2009)
  [arXiv:0805.2944 [hep-ph]].
}

\lref\SunRtree{
  Z.~Sun,
  ``Tree level Spontaneous R-symmetry breaking in O'Raifeartaigh models,''
  JHEP {\bf 0901}, 002 (2009)
  [arXiv:0810.0477 [hep-th]].
}

\lref\Fbreak{L.~O'Raifeartaigh,
``Spontaneous Symmetry Breaking For Chiral Scalar Superfields,''
Nucl.\ Phys.\ B{\bf 96}, 331 (1975).
}

\lref\WessCP{
  J.~Wess and J.~Bagger,
  ``Supersymmetry and supergravity,''
{\it  Princeton, USA: Univ. Pr. (1992) 259 p}
}

\lref\BaggerGM{
  J.~A.~Bagger and A.~F.~Falk,
  ``Decoupling and Destabilizing in Spontaneously Broken Supersymmetry,''
  Phys.\ Rev.\  D {\bf 76}, 105026 (2007)
  [arXiv:0708.3364 [hep-ph]].
}

\lref\RocekNB{
  M.~Rocek,
  ``Linearizing The Volkov-Akulov Model,''
  Phys.\ Rev.\ Lett.\  {\bf 41}, 451 (1978).
}

\lref\GirardelloWZ{
  L.~Girardello and M.~T.~Grisaru,
  ``Soft Breaking Of Supersymmetry,''
  Nucl.\ Phys.\  B {\bf 194}, 65 (1982).
}

\lref\shih{
  D.~Shih,
  ``Spontaneous R-symmetry breaking in O'Raifeartaigh models,''
  JHEP {\bf 0802}, 091 (2008)
  [arXiv:hep-th/0703196].
  }

\lref\IvanovMY{
  E.~A.~Ivanov and A.~A.~Kapustnikov,
  ``Relation Between Linear And Nonlinear Realizations Of Supersymmetry,''
  JINR-E2-10765, Jun 1977
}

\lref\IvanovMX{
  E.~A.~Ivanov and A.~A.~Kapustnikov,
  ``General Relationship Between Linear And Nonlinear Realizations Of
  Supersymmetry,''
  J.\ Phys.\ A  {\bf 11}, 2375 (1978).
}
\lref\IvanovBPA{
  E.~A.~Ivanov and A.~A.~Kapustnikov,
  ``The Nonlinear Realization Structure Of Models With Spontaneously Broken
  Supersymmetry,''
  J.\ Phys.\ G {\bf 8}, 167 (1982).
}

\lref\VolkovIX{
  D.~V.~Volkov and V.~P.~Akulov,
  ``Is the Neutrino a Goldstone Particle?,''
  Phys.\ Lett.\  B {\bf 46}, 109 (1973).
}

\lref\AffleckVC{
  I.~Affleck, M.~Dine and N.~Seiberg,
  ``Dynamical Supersymmetry Breaking In Chiral Theories,''
  Phys.\ Lett.\  B {\bf 137}, 187 (1984).
}

\lref\SamuelUH{
  S.~Samuel and J.~Wess,
  ``A Superfield Formulation Of The Nonlinear Realization Of Supersymmetry And
  Its Coupling To Supergravity,''
  Nucl.\ Phys.\  B {\bf 221}, 153 (1983).
}

\lref\LindstromKQ{
  U.~Lindstrom and M.~Rocek,
  ``Constrained Local Superfields,''
  Phys.\ Rev.\  D {\bf 19}, 2300 (1979).
}

\lref\BaggerHH{
  J.~Bagger, E.~Poppitz and L.~Randall,
  ``The R axion from dynamical supersymmetry breaking,''
  Nucl.\ Phys.\  B {\bf 426}, 3 (1994)
  [arXiv:hep-ph/9405345].
}

\lref\DineSW{
  M.~Dine and J.~Kehayias,
  ``Discrete R Symmetries and Low Energy Supersymmetry,''
  arXiv:0909.1615 [hep-ph].
}

\lref\AdamsHP{
  A.~Adams, A.~Jenkins and D.~O'Connell,
  ``Signs of analyticity in fermion scattering,''
  arXiv:0802.4081 [hep-ph].
}

\lref\LuoIB{
  H.~Luo, M.~Luo and S.~Zheng,
  ``Constrained Superfields and Standard Realization of Nonlinear
  Supersymmetry,''
  arXiv:0910.2110 [hep-th].
}


\rightline{ hep-th/yymmnnn}
\rightline{SCIPP-2009/14}

\Title{
} {\vbox{\centerline{A Bound on the Superpotential}
}}
\medskip

\centerline{\it Michael Dine${\,^{(a)}}$, Guido Festuccia${\,^{(a)}}$ and Zohar Komargodski${\,^{(b)}}$}
\bigskip
\centerline{${^{(a)}}$ Santa Cruz Institute for Particle Physics}
\centerline{University of California Santa Cruz}
\centerline{1156 High street, Santa Cruz, CA 95064}
\bigskip
\centerline{${^{(b)}}$ School of Natural Sciences}
\centerline{Institute for Advanced Study}
\centerline{Einstein Drive, Princeton, NJ 08540}

\smallskip

\vglue .3cm

\bigskip
\noindent We prove a general bound on the superpotential in theories with broken supersymmetry and broken
$R$-symmetry, $\vert\langle W \rangle\vert < {1 \over 2} f_a F$, where $f_a$ and $F$ are the $R$-axion and
Goldstino decay constants, respectively. The bound holds for weakly coupled as well as strongly coupled
theories, thereby providing an exact result in theories with broken supersymmetry. We  briefly discuss several possible
applications.

 \Date{10/2009}


\newsec{Introduction}

In models with rigid supersymmetry, $R$-symmetry is intimately tied to questions of supersymmetry breaking.  The
theorem of Nelson and Seiberg~\refs{\nelsonseiberg} asserts that a calculable (and generic) supersymmetric
theory exhibits supersymmetry breaking only if it possesses an $R$-symmetry, and that a generic calculable
theory with broken $R$-symmetry breaks supersymmetry.  Metastable breaking, one expects, will usually require an
approximate $R$-symmetry~\IntriligatorPY.

In a supersymmetric and $R$-symmetric theory, one is not allowed to
add an arbitrary constant to the superpotential. It can only be
generated dynamically in the IR upon $R$-symmetry breaking. Thus, in
$R$-symmetric theories, a candidate order parameter for $R$-symmetry
breaking is the superpotential itself. As we will see, in $R$ symmetric theories,
$\langle W \rangle$ is indeed a good order parameter, as it is directly measurable
in low energy scattering experiments.  It is therefore natural to ask what
is the relation of the superpotential VEV to other order parameters
in the problem, such as the $R$-axion decay constant and the vacuum
energy (which is related to the Goldstino decay constant).  Understanding $\langle W \rangle$
 could elucidate the connection between $R$-symmetry breaking and SUSY breaking~\nelsonseiberg.

An additional circumstance in which $\langle W \rangle$ takes on physical significance
is in coupling a theory to supergravity.
There, $\langle W \rangle$ plays an important role in accommodating
a small cosmological constant, and is directly related to the value of the
gravitino mass and the mass of any would-be $R$-axion.

The simplest model one might consider is a free theory of a single
chiral superfield $Z$ with $R(Z)=2$ \eqn\suplinear{W = f Z~,\qquad
K= Z^\dagger Z~.} Both the boson and the fermion in $Z$ are
massless, and there is a moduli space of SUSY-breaking vacua
parameterized by the expectation value of the bottom component of
$Z$, $\langle z\rangle $. Writing $z=\vert z\vert e^{2ia}$, we see that the $R$-axion decay constant\foot{We define the $R$-axion decay constant, $f_a$, as the coefficient of the kinetic term of the axion $$-f_a^2(\partial a)^2~.$$.} is $f_a = 2 \vert
\langle z\rangle \vert$ while $\langle W \rangle = f \langle
z\rangle$ and the vacuum energy density is $\vert f\vert^2$. So,
\eqn\toybound{\vert\langle W \rangle\vert = {1 \over 2} f_a F~,} where $F$ is the Goldstino decay constant, given by the vacuum energy
 $F^2=\vert f\vert^2$.

 This trivial model is illuminating. In this paper,
we will prove an inequality, which holds in {\it any} theory with
broken supersymmetry and broken $R$-symmetry: \eqn\inequality{\vert
\langle W \rangle \vert \le {1 \over 2} f_a F~.}
We will also argue that interacting theories always lead to a strict inequality
\eqn\inequalityi{\vert
\langle W \rangle \vert < {1 \over 2} f_a F~.}

We will prove the result in a sequence of situations. First, in
section 2,  we will consider general O'Raifeataigh-like models, at
tree-level. Here the proof is quite simple. We will illustrate the
theorem with models that saturate the bound at tree-level, and
which do not, and understand the distinction. We will also consider
renormalizable theories with gauge interactions, for which $D$-terms
may be non-zero, and show that the inequality remains true.

In section~3, we consider linear sigma models, i.e. theories with
chiral fields and a general K\"ahler potential and superpotential.
At the level of two derivative terms in the effective action, the theorem
is readily proven for these theories as well.  However, for a class of theories which
saturate the bound, it is necessary to look at higher orders in the derivative
expansion.  In this case, the proof of the theorem invokes considerations
of unitarity along the lines of ~\refs{\nima}.

In section~4 we address the most general case, which need not be a
calculable theory. To this end we use the machinery of non-linear
effective Lagrangians, developed recently in~\refs{\nonlinear}.
After reviewing the necessary background, we show that the desired
inequality translates into an inequality between various parameters
in the effective non-linear Lagrangian. Along with standard
manipulations in such effective theories, we use the consistency
conditions for effective field theories discussed in~\refs{\nima} to
prove that non-trivial interacting theories always satisfy~\inequalityi.

A number of technical details are covered in an appendix.

\newsec{The Bound in (Gauged) O'Raifeartaigh-Like Models}

In this section we will prove the bound in theories with canonical
K\"ahler potential where both SUSY and the $R$-symmetry are broken.
We will start our analysis from the case of O'Raifeartaigh-like
models. We then briefly comment on the extension to models with
gauge fields.

The general model with canonical K\"ahler potential and no gauge
interactions takes the form \eqn\ORaif{\CL=\int d^2\theta d^2 \bar
\theta \;\sum_i\bar\Phi_i \Phi_i+\int d^2 \theta\, W(\Phi_i)+c.c.~.}
The chiral fields $\Phi_i$ have charge $q_i$ under the $R$-symmetry.
By assumption $W(\Phi_i)$ is a holomorphic function of the $\Phi_i$
of $R$-charge $2$. Before proving our inequality in these models, we
would like to outline rather general features that are helpful to
develop some intuition.

Let us consider theories of the form~\ORaif\ that have a
SUSY-breaking and $R$-symmetry breaking vacuum at some point
$\phi_i^{(0)}$, where $\phi_i$ is the bottom component of $\Phi_i$.
The spectrum therefore comprises a massless Goldstino and a massless
$R$-axion. As discussed in~\RayWK, classically, there is always a
complex massless boson (in fact a whole flat direction) generated by
the transformation \eqn\flatdirection{\phi_i^{(0)}\rightarrow
\phi_i^{(0)}+\alpha \left({\partial W\over \partial\phi_i}\right)^*~,} for any complex $\alpha$. A simple way
to remember this massless direction is as the bosonic superpartner
of the Goldstino~\KomargodskiJF.

One can distinguish two cases:

{\item{$\bullet$} If the $R$-symmetry is broken everywhere on the pseudomoduli space~\flatdirection, the
Goldstino superpartner is linearly independent of the $R$-axion~\KomargodskiJF. As a consequence, generically,
the pseudomoduli space is three real dimensional.}

{\item{$\bullet$} If somewhere on the pseudomoduli
space~\flatdirection\ the $R$-symmetry is restored then often
these are the only two real flat directions.
The $R$-axion is embedded into this complex flat direction as some
phase coordinate around the $R$-restoring point.
}

\noindent

We will see that the first case satisfies the inequality $\vert
\langle W \rangle \vert < {1 \over 2} f_a F$ while, as in the second
case, if the $R$-axion direction is embedded in~\flatdirection\ we get
$\vert \langle W \rangle \vert = {1 \over 2} f_a F$.

\subsec{Proof of the Bound for O'Raifeartaigh-Like Models}
Since, by assumption, $W(\Phi_i)$ is a holomorphic function of the
$\Phi_i$ of $R$-charge $2$ we have the following identity ($q_i$ stands for the $R$-charge of $\Phi_i$)
\eqn\WRsym{2W(\Phi_i)=\sum_j q_j \Phi_j{\partial W(\Phi_i)\over
\partial \Phi_j}~.}
We now consider the bottom component of~\WRsym\ and define two
complex vectors $w_i= q_i \phi_i$ and $F_i^{\dagger}= {\partial
W\over
\partial \phi_i}$. The Goldstino decay constant $F$ is given by\foot{The inner product $\langle , \rangle$ is the standard (Hermitian) Euclidian inner product.} \eqn\Golddecay{F^2=\sum_j
F_j^{\dagger}F_j=\langle {\bf\nabla W},{\bf\nabla W}\rangle~.}

The scalar fields are parameterized by the $R$-axion $a$ as
$\phi_i=|\phi_i|e^{iq_ia}$. The kinetic part of the Lagrangian for
$a$ is: \eqn\norax{ -\partial_{\mu} a
\partial^{\mu} a \sum_i q_i^2 |\phi_i|^2 } from which we read the
$R$-axion decay constant $f_a^2=\sum_i q_i^2 |\phi_i^{(0)}|^2
=\langle {\bf w},{\bf w}\rangle$.

We use the Cauchy-Schwarz inequality in~\WRsym\ and obtain
\eqn\Schwarz{ 4|W|^2=|\langle {\bf F},{\bf w}\rangle|^2\leq \langle {\bf w},{\bf w}\rangle
\langle {\bf \nabla W},{\bf\nabla W} \rangle.} This can be rewritten as $2
|\langle W\rangle|\leq  f_a F$, establishing the bound.

An immediate corollary is that we can classify the models which
saturate the bound. $R$-symmetry transformations are generated by
\eqn\Rtrans{\phi_i^{(0)}\rightarrow \phi^{(0)}_i+i\epsilon
q_i\phi^{(0)}_i~.} From~\Schwarz\ we see that the bound is saturated
if and only if the vector $w_i=q_i\phi_i^{(0)}$ is proportional to
$F_i=W_i^*$. This means that the $R$-symmetry transformation~\Rtrans\
is part of the canonical pseudomodulus space~\flatdirection.
Theories in which the canonical pseudomoduli
space~\flatdirection\ does not contain an $R$-symmetric point cannot
saturate the bound.

\subsec{Examples}

As a first example consider the original O'Raifeartaigh model~\refs{\Fbreak} with superpotential
\eqn\ORex{W=X\left({\lambda\over 2} A^2-f\right)- m  YA~.} where $X$ and $Y$ have $R$-charge $2$ while $A$ is
neutral. This model is generic under the assumption of an extra $Z_2$ symmetry changing the sign of $A$ and $Y$.
For $m^2\geq\lambda f$ the lowest lying pseudomodulus space of vacua is given by $Y=A=0$. The only field with a
nonzero $F$-term is $X$ and  $F^{\dagger}_X= f$.  On this branch $W_{eff}=-f X $. The $R$-axion is embedded in
the phase of $X$ with $f_a=2|X|$. Then $f_aF=2 |X| |f|=2|W|$ and the bound is saturated. The story for the other
branch (which is stable for $m^2\leq\lambda f$) of the O'Raifeartaigh model~\ORex\ is very similar.

A model where the $R$-symmetry is broken everywhere on the pseudomoduli space was given in~\refs{\ggmimpl}; this
class of models was considered in more generality in~\refs{\KomargodskiJF} and~\refs{\SunRtree}. As an example
consider the following superpotential:
 \eqn\treebreak{ W= X \left( \gamma \phi_{2/3} \phi_{-2/3}
-\mu^2\right)+{\delta\over 3} \phi_{2/3}^2 \chi_{2/3} +m_1 \phi_{2/3} Y_{4/3} +m_2 \phi_{-2/3} Z_{8/3} + \lambda
\chi_{2/3}^3~.} The subscripts are the $R$-charges of the various fields. $X$ has $R$-charge 2. This
superpotential is the most general compatible with a $Z_2$ symmetry under which all fields but $X$ and
$\chi_{2/3}$ are odd. When $\sigma$, defined by $\mu^2={m_1 m_2\over \gamma}(1+\sigma^2 \gamma^2)$, is real
there is a stable branch of the pseudomoduli space where the $R$-symmetry is everywhere broken: \eqn\branch{
\phi_{2\over 3}=m_2\sigma e^{i\theta}~,\qquad \phi_{-{2\over 3}} = {m_1 \sigma}e^{-i \theta}~,\qquad
\chi_{2\over 3}=i m_2{1\over 3} \sqrt{\delta \over \lambda}\sigma e^{i\theta}~, } while the fields with nonzero
$F$-terms are related by: \eqn\pseX{Z_{8\over 3}=-\gamma\sigma X e^{i \theta}~,\qquad Y_{4\over 3}=Z_{8\over
3}e^{-2 i \theta}-{2 m_2^2 \delta^{3\over 2} \over 3 m_1 \lambda^{1\over 2}}\sigma e^{2 i \theta}~.} We see that
the pseudomoduli space is three real dimensional, in accord with the fact that the $R$-symmetry is nowhere
restored.

On this branch the total vacuum energy (and hence the Goldstino decay constant) is \eqn\vacuumenergy{
F^2={m_1^2 m_2^2\over \gamma^2}\left(1+2\sigma^2 \gamma^2\right)\leq \mu^4~.}
 The $R$-axion is proportional to $\theta(x)$. It
can checked explicitly that $2|W|<f_a F$ everywhere on the pseudomoduli space for $\sigma\neq 0$. For
$\sigma=0$ the $R$-symmetry is restored at $X=0$ and the bound is saturated. The resulting expressions are
cumbersome in general, therefore, we just quote the result at leading order for large $|X|$ \eqn\leadingX{f_a^2
F^2-4|W|^2={8\over 9} m_1^2 m_2^2 \sigma^2 (1+2 \gamma^2\sigma^2)|X|^2+O\left(X^0\right)\geq0 ~.}

\subsec{Adding $D$-Terms}

The discussion above is readily generalized to include possible
$D$-terms. Indeed, we can repeat the discussion around eqns.~\WRsym\
and \norax. The only change is that the Goldstino decay constant,
$F$, receives additional contributions, \eqn\dtermcontribution{
F^2  = \sum_i \biggl\vert {\partial W \over \partial \phi_i}
\biggr\vert^2 + \sum_a D_a^2~.} So $ F^2
> \langle {\bf \nabla W},{\bf \nabla W} \rangle$, strengthening the bound.

\newsec{The Bound in Sigma Models}

In this section, we show that the bound is satisfied for theories
described by arbitrary superpotential and  K\"ahler potential
(consistent with an $R$-symmetry)
\eqn\theoryi{K(\phi_i,\bar\phi_j)~,\qquad W(\phi_i)~.} Let us denote
for simplicity \eqn\definition{{\cal M}_{l\bar m}={\partial^2 K\over
\partial {\phi_l}\partial {\bar\phi_m}}~.} The matrix $g={\cal M}+{\cal M}^\dagger$ is Hermitian.
It is also positive definite around the configurations in field
space we are interested in, since otherwise there are ghosts.
Therefore, we can decompose
\eqn\decomposition{g=LL^\dagger~,\qquad
g^{-1}=(L^\dagger)^{-1}L^{-1}~.} Even when the K\"ahler potential
is non-canonical,~\WRsym\ still holds and we can rewrite (by
inserting the identity matrix\foot{Note that the index $j$ takes values in both barred and unbarred indices, while $i$,$k$ only take values in the unbarred indices. We hope our attempt not to clutter the notation will not cause a confusion. })
\eqn\WRSymi{2W=\sum_iq_i\phi_i{\partial W\over
\partial \phi_i}=\sum_{i,j,k}q_i\phi_iL_{ij}L^{-1}_{jk}{\partial
W\over
\partial \phi_k}~.} It is useful to define the two vectors $w_j=\sum_iq_i\phi_iL_{ij}$,
$v_j^{\dagger}=\sum_kL^{-1}_{jk}{\partial W\over \partial \phi_k}$.
The Goldstino decay constant, $F$, is related to the vacuum energy
which is just $F^2=\langle {\bf v},{\bf v} \rangle $, while the $R$-axion decay
constant can be read off the kinetic terms of the sigma
model~\theoryi. The kinetic terms contain the matrix $g$ and
therefore the decay constant satisfies $f_a^2 =\langle {\bf w},{\bf w}\rangle$.

Using the Cauchy-Schwarz inequality again we get \eqn\Schwarzi{
4|W|^2=|\langle {\bf v},{\bf w}\rangle|^2\leq \langle {\bf w},{\bf w}\rangle \langle {\bf v},{\bf v}
\rangle=f_a^2F^2~,} or in other words
 \eqn\Schwarzii{
2|W|\le f_aF~,} which confirms the bound we claim.

\subsec{An Overture to Beyond the Sigma Model}

While very suggestive, however, the tree-level argument in the
general sigma model does not prove the theorem in complete generality.  There are two
limitations. First, while many models of dynamical supersymmetry
breaking (in particular models where SUSY is broken at tree-level)
permit a low energy description of SUSY breaking in terms of a field
theory with linearly realized supersymmetry, there are many models
which do not. Perhaps the most well known examples are the $SU(5)$
and $SO(10)$ models of~\AffleckVC,\AffleckMF. Secondly, we have only
considered tree-level sigma models, forbidding derivative
corrections in the K\"ahler potential and (not dissimilar) radiative
corrections.

To show that derivative corrections in the K\"ahler potential may play a decisive role, consider the following
tree-level theory of a single chiral superfield $Z$ with $R$-charge~2 \eqn\onefield{K = K(ZZ^\dagger)~,\qquad W
= f Z~.} Such models arise as an effective description of theories with a characteristic mass
scale $M$ and a parameterically small SUSY breaking scale $\vert f \vert \ll M^2$. This separation of scales
guarantees that one should be able to describe the theory at low energies in terms of an effective action with
linearly realized supersymmetry.

Repeating our analysis above in this simple case, one sees that the bound is saturated (since the $R$-symmetry
breaking must occur, if at all, in the same direction as SUSY breaking). The description~\onefield\ is a valid
effective description up to two derivatives. But generically, there are also higher derivative corrections.

At the level of terms with two derivatives in superspace, the effective action may include various contributions. Let us study, for example, the term
\eqn\fourderivative{-\int d^4 \theta {\epsilon \over M^4} Z^\dagger
Z\partial_\mu Z^\dagger \partial^\mu Z~.}
This term changes $f_a$ without affecting the Goldstino decay constant. Indeed since it has explicit derivatives
it has no effect on the vacuum energy, but it does change the normalization of the kinetic term for~$z$. In addition, this term does not change the superpotential VEV $\langle W\rangle $.\foot{ In the next section we will define the observable $\langle W\rangle$ more carefully.}

Since the original theory~\onefield\ always saturates the bound, a negative sign for $\epsilon$ would seem to
contradict the inequality: the change in $f_a$ in the presence of~\fourderivative\ is such that $f_a$
decreases if $\epsilon$ is negative. Therefore, some principle should dictate that $\epsilon$ is positive if the
inequality is true.

This is the first point where we see the importance of unitarity arguments in effective field theory. The operator~\fourderivative\ can be shown to arise with a definite sign of $\epsilon$ in effective theories
that have UV completions. In the rest of this subsection we will show that this is the case by carefully
studying the theory~\onefield\ deformed by~\fourderivative.

We define the $R$-axion as usual to be the phase of $z$, $z=|z|e^{2ia}$. Suppose for a moment that $\epsilon=0$.
The Lagrangian, after integrating out the radial mode $|z|$, contains the usual kinetic terms for the axion and
Goldstino, but it also contains some interaction terms between the axion and the Goldstino. As we show in
appendix~$\aleph$, the leading interaction is of the
form~\eqn\leadingint{\CL_{interaction}\approx c_1\psi^2\partial_\mu a\partial^\mu a+c.c.~.} Another possible coupling one
could imagine has three derivatives and takes the form
$i\left(\partial^\nu\psi\sigma^\mu\bar\psi\right)\partial_\mu a\partial_\nu a$, however, as we show explicitly in
appendix~$\aleph$, it does not arise at tree-level in the theory with $\epsilon=0$.

Introducing the deformation~\fourderivative\ induces this new coupling of two fermions to two axions, with
coefficient proportional to $\epsilon$. We will now argue that, very similarly to examples discussed in~\nima,
the theory \eqn\theory{\CL=-f_a^2(\partial_\mu
a)^2+i\partial_\mu\bar\psi\bar\sigma^\mu\psi+\left(c_1\psi^2(\partial_\mu a)^2+c.c.\right)
-ic_2\left(\partial^\nu\psi\sigma^\mu\bar\psi\right)\partial_\mu a\partial_\nu a\cdots~,} where $\cdots$ stand
for self interactions of the axion and the Goldstino that we are not interested in, has superluminal modes if
$c_2$ has the wrong sign.

Let us look at the propagation of the Goldstino $\psi$ in a background with $\partial_{\mu}a=V_{\mu}$, where
$V_{\mu}$ is a constant four vector. The Lagrangian for the Goldstino becomes
\eqn\theory{\CL=i\partial_\mu\bar\psi\bar\sigma^\mu\psi+\left(c_1V^2\psi^2+c.c.\right) -ic_2V_\mu
V_\nu\partial^\nu\psi\sigma^\mu\bar\psi+\cdots~.}
We are expanding in small $V$, as appropriate in effective field theory. Note that around the background $\partial_{\mu}a=V_{\mu}$ the fermion $\psi$ is massive with mass $\sim V^2$. As usual, in the final dispersion relation the mass appears squared, so it can be dropped. The remaining terms yield the following equation of motion in momentum space
\eqn\eom{\left(k_\mu\bar\sigma^\mu-c_2(V\cdot k)V_\mu\bar\sigma^\mu\right)\psi=0~.}
Let us multiply this equation by $V_\rho\sigma^\rho$ from the left.  We see that at leading order in $V$
\eqn\eomi{V_\rho k_\mu \sigma^\rho \bar\sigma^\mu \psi=0~.}
The next step is to multiply~\eom\ by $k_\rho\sigma^\rho$ from the left to obtain
\eqn\eomii{\left(-k^2\delta^{\alpha}_\beta-c_2(V\cdot k)k_\rho V_\mu\left(\sigma^\rho\bar\sigma^\mu\right)_\beta^\alpha\right)\psi_\alpha=0~.}
We can symmetrize the second term using~\eomi. We obtain
\eqn\eomii{\left(k^2-2c_2(V\cdot k)^2\right)\psi_\beta=0~.}
We see that the dispersion relation is corrected by $(V\cdot k)^2$, which has a definite sign. This leads to superluminal modes unless $c_2\geq 0$.

Expanding the term~\fourderivative\ in terms of component fields, we find that it leads to an operator of the
form $c_2$ with coefficient $c_2=8\epsilon v^2/M^4$, where $v$ is the VEV of $|z|$. If the operator of
eqn.~\fourderivative\ were the only four derivative operator, we see that unitarity would require the
coefficient $\epsilon$  to be positive.\foot{The connection between superluminal modes and unitarity bounds was
demonstrated in~\nima. These ideas apply in our context equally well. In particular, while the absence of
superluminal modes only suggests that $c_2\geq 0$, unitarity gives $c_2>0$ because it relates $c_2$ to an
integral of a total cross section. } As we commented above, a positive $\epsilon$ increases the decay constant
of the $R$-axion, leaving the other quantities intact, therefore, we would find, for any non-zero $\epsilon$,
\eqn\boundi{ 2|W|<f_aF~.} The operator of eqn. $\fourderivative$ is just one of several operators which can
potentially correct $f_a$.  In order to provide a general proof of the bound, we will see that it is necessary
to consider the effective theory at very low energies.

In spite of the fact that at tree-level the bound can  sometimes be saturated, we will show that nontrivial
interacting theories will always satisfy~\boundi. It is not difficult to verify that in simple perturbative
models, such as those of~\shih, the bound is indeed satisfied at one-loop.

\newsec{A Proof of the Bound via Low Energy Effective Field Theory}

In the previous two sections we have seen how the bound arises in tractable field theories and we have also witnessed the importance of unitarity arguments in effective field theory. The arguments of the previous sections cover a variety of models, but not the interesting, and potentially important ones where the scale of supersymmetry breaking is not
 small compared to other characteristic mass scales of the theory, so that SUSY breaking cannot be described by a superpotential with linearly realized SUSY. Models of the latter kind are pervasive; examples include the $SU(5)$
and $SO(10)$ models of~\AffleckVC,\AffleckMF.
In addition, while in the previous section we discussed one particular derivative correction characteristic
of perturbative models, we would like to be able to control all of them along with all the possible  radiative corrections.

Therefore, our goal in this section is to extend the validity of our bound to these cases as well. All we know is that these models break supersymmetry and have a spontaneously broken $R$-symmetry. Therefore, the low energy spectrum consists of an $R$-axion and a Goldstino. However, this cannot be an arbitrary effective theory, rather, it has to be constrained by nonlinearly realized SUSY and $R$-symmetry.

As explained in~\nonlinear, such nonlinear theories are most conveniently organized in terms of a set of superfields that satisfy algebraic constraints. Below we review this formalism, focusing on the aspects pertinent to this work. After reviewing the necessary ingredients, we will derive the bound   \eqn\boundii{
2|W|<f_aF~,}
establishing its validity as a nonperturbative result.

\subsec{A Review of Nonlinear SUSY and $R$-Symmetry}

The low energy effective theory we are after includes the Goldstino $G_\alpha(x)$ and the $R$-axion $a(x)$. The Lagrangian should respect nonlinearly realized supersymmetry as well as nonlinearly realized $R$-symmetry. The latter is easier to understand, as it includes an inhomogeneous shift of the $R$-axion accompanied by a rotation of the Goldstino field and the superspace coordinate
\eqn\nonlinearR{a\rightarrow a+\xi~,\qquad G_\alpha\rightarrow e^{i\xi}G_\alpha~,\qquad \theta_\alpha\rightarrow e^{i\xi}\theta_\alpha~.}
On the other hand, the action of nonlinear supersymmetry is more complicated since the underlying group structure is non-Abelian. There are many approaches to nonlinear realizations of supersymmetry, see e.g.~\SamuelUH\ (more references can be found in~\nonlinear).
In this work we will adopt the conventions and approach of~\nonlinear\ as it easily allows to describe off-shell effective actions that may or may not include particles in addition to the Goldstino.\foot{Some comments on the relation between~\nonlinear\ and previous approaches can be found in~\LuoIB.}

The main point is that we will use the power of supersymmetry and superfields, but our superfields will satisfy some constraints. The effect of these constraints, as we will momentarily see, is to remove some degrees of freedom from the conventional superfields.

To see how this works, we start from a chiral superfield, $X_{NL}$, that satisfies the constraint
\eqn\consX{X_{NL}^2=0~.}
This constraint eliminates the complex boson in the bottom component of $X_{NL}$ but leaves the fermion component as well as an auxiliary field. The solution to~\consX\ is
\eqn\conssol{X_{NL}={G^2\over 2F_X}+\sqrt{2} \theta G+\theta^2 F_X~,}
where all the variables are functions of $y^{\mu}=x^{\mu}+i\theta \sigma^{\mu} \bar \theta$.
It is therefore natural to identify the fermion surviving the constraint~\consX\ as the Goldstino.

If the Goldstino is the only massless particle, we can easily write effective Lagrangians by using superspace and the superfield $X_{NL}$. These by construction respect nonlinear SUSY.
The simplest possible Lagrangian is
\eqn\AV{\CL=\int d^4\theta X_{NL}^\dagger X_{NL}+\left(\int d^2\theta fX_{NL}+c.c.\right)~.}
This is no other than the Akulov-Volkov theory~\VolkovIX\ , that in components includes the following terms.\foot{For alternative descriptions of this theory see~\refs{\IvanovMY\IvanovMX\RocekNB\LindstromKQ-\IvanovBPA}.}
\eqn\AVcomp{{\cal L}=-f^2+i\partial_\mu\bar G \bar\sigma^\mu G +{1\over 4f^2}\bar G ^2\dal G ^2+\cdots ~,}
where $\dots$ stand for terms with more Goldstinos.

While the constraint~\consX\ forbids a nontrivial K\"ahler potential, corrections to~\AV\ with derivatives are allowed (and generally appear from microscopic models). The natural way to control these corrections is to assign the Goldstino $G_\alpha$ effective scaling dimension $-1/2$ and therefore the superfield $X_{NL}$ has scaling dimension $-1$
\eqn\scalingX{S(X)=-1~.}
This choice forces us to assign $d\theta$ scaling dimension $+1/2$. The terms in the Lagrangian~\AV\  then have scaling dimension zero. It can be proven~\nonlinear\ that at scaling zero the Lagrangian~\AV\ is the most general possible up to field redefinitions. The theory~\AV\ has only one free parameter, the SUSY breaking scale.

To include an $R$-axion (or more generally a Goldstone boson) we introduce a chiral superfield $\CA_{NL}$, satisfying the following constraint
\eqn\Raxion{X_{NL}\left(\CA_{NL}-\CA_{NL}^\dagger\right)=0~.}
Out of all the degrees of freedom in a conventional chiral superfield, the constraint~\Raxion\ leaves only one real degree of freedom (and no auxiliary fields). In components, the superfield $\CA_{NL}$ takes the form
\eqn\Rsuper{\eqalign{
{\cal A}_{NL}= & H+i\sqrt2\theta\sigma^\mu\left(\bar G\over \bar F_X\right)\partial_\mu H+\theta^2\left(-\partial_\nu\left({\bar G\over \bar F_X}\right)\bar\sigma^\mu\sigma^{\nu}{\bar G\over \bar F_X}\partial_\mu H+{1\over 2\bar F_X^2}\bar G^2\dal H\right) ~,}}
where
\eqn\Haxion{H=a+{i\over 2} ({G\over F_X}\sigma^\mu{\bar G\over \bar F_X})\partial_\mu a+\cdots~~,}
where the ellipses  stand for terms with more fermions and derivatives. We will not need them here.\foot{The corrections are easily derived from~\Raxion. An explicit expression is given in~\nonlinear.}

We see that $R$-symmetry acts on $\CA_{NL}$ by shifts $\CA_{NL}\rightarrow\CA_{NL}+\xi$. Of course, the shift preserves the constraint~\Raxion. Due to this action of $R$-symmetry it is natural to define an exponentiated superfield
\eqn\Rsup{\CR_{NL}=e^{i\CA_{NL}}~,}
which under $R$-symmetry transforms as $\CR_{NL}\rightarrow e^{i\xi}\CR_{NL}$. The constraint~\Raxion\ becomes
\eqn\Raxioni{X_{NL}\left(\CR^\dagger_{NL} \CR_{NL}-1\right)=0~.}

We are now ready to write Lagrangians. As we are going to use superspace and superfields, both nonlinear supersymmetry and nonlinear $R$-symmetry will be manifest. Following the same idea as around~\AV\ we get the following Lagrangian
\eqn\Lagaxgol{\CL_{Goldstino-axion}= \int d^4\theta\left(|X_{NL}|^2 + f_a^2|\CR_{NL}|^2\right)+\int d^2\theta \left(fX_{NL}+\tilde f\CR_{NL}^2\right)+c.c. ~.}
The natural scaling dimension of the $R$-axion is zero and so
\eqn\scalingR{S(\CR_{NL})=0~.}

We see that the effective Lagrangian contains three independent parameters. $f$ and $f_a$ are identified as the SUSY breaking scale and the $R$-axion decay constant, respectively. The remaining parameter $\tilde f$ corresponds to the VEV of the superpotential.

Similarly to the case in~\AV, it can be shown that in an expansion in derivatives (more precisely, in the scaling $S$) the theory~\Lagaxgol\ is the leading universal theory at low energies (up to field redefinitions that can be absorbed in redefinitions of $f,f_a,\tilde f$).

Let us pause for a moment to discuss the parameter $\tilde f$. In theories that break supersymmetry, the VEV of the superpotential is, in general, not a holomorphic function of the superpotential couplings. Furthermore, in the presence of covariant derivatives its naive definition is ambiguous.\foot{For example, because we can always add to the superpotential terms like $\bar D^2 \CO^\dagger$ with $\CO^\dagger$ anti-chiral and $R$-neutral. This does not change the physical theory but may affect the value of $\langle W\rangle$. } The parameter $\tilde f$ provides a precise physical definition of $\langle W\rangle$, which generalizes what one would naively call the VEV of the superpotential in simple theories like those analyzed in the previous sections.

We conclude that if we wish to analyze the interactions between an $R$-axion and the Goldstino at very low energies, we should study the theory~\Lagaxgol\ and substitute the component expressions~\conssol,\Rsuper,\Haxion\ for the superfields $X_{NL}$, $\CR_{NL}$. This is done in detail in the next subsection, where we provide the most general proof of the bound.

\subsec{General Proof of the Bound}

In this section we will describe the interactions resulting from~\Lagaxgol.  We will only keep terms with at
most four fields, as this will suffice for the argument we are about to make. We directly substitute the
expressions~\conssol,\Rsuper,\Haxion\ into~\Lagaxgol\ and get the following
\eqn\Lagcomp{\eqalign{&\CL_{Goldstino-axion}=-f^2+i\partial_\mu\bar G \bar\sigma^\mu G-f_a^2(\partial a)^2
+{1\over 4f^2}\bar G ^2\dal G ^2-2i{f_a^2\over f^2} \partial^{\mu}G\sigma^\lambda \bar{G} \partial_{\mu}a
\partial_{\lambda}a\cr&+\left( {\tilde f\over f^2}\bar G\bar\sigma^\mu\sigma^\nu\partial_\nu \bar
G\partial_\mu\left(e^{2ia}\right) -{i\tilde f\over f^2}\bar G^2e^{2ia}\sqr{6}{6}\ a+{2\tilde f\over f^2}\bar
G^2(\partial a)^2+c.c.\right)~.}} We have dropped quartic terms that are proportional to the free equations of
motion, as they effectively contain more than four fields. Note that the first two terms in the second line are
proportional to the free equations of motion of the Goldstino and the axion, respectively. {\it However,  the
key point is that these are cubic terms, so they can lead to effective quartic operators.}

To calculate the quartic terms we may use field redefinitions to eliminate the cubic terms. Let us define a new
Goldstino field $\tilde G_\alpha$ and a new axion $\tilde a$ related to the variables in~\Lagcomp\ via
\eqn\redef{\tilde G=G+i{\tilde f\over f^2}\bar G\bar\sigma^\mu\partial_\mu\left(e^{2ia}\right)~,\qquad \tilde
a=a-i{\tilde f\over 2f^2f_a^2}e^{2ia}\bar{\tilde G}^2+c.c.~. } Note that these new Goldstino and axion fields
are not the same as those sitting in the superfields, $X_{NL}$, $\CR_{NL}$. For example, their transformation
laws under supersymmetry are different. The Lagrangian~\Lagcomp\ can be written in terms of the new
fields~\redef. The result (again, up to terms with more than four fields) is
\eqn\Lagcompi{\eqalign{&\CL_{Goldstino-axion}=-f^2+i\partial_\mu\bar {\tilde G} \bar\sigma^\mu \tilde
G-f_a^2(\partial a)^2 +\left({1\over 4f^2}-{|\tilde f|^2\over 2f^4f_a^2}\right)\bar {\tilde G} ^2\dal
\tilde G ^2\cr&-2i\left({f_a^2\over f^2}-4{|\tilde f|^2\over f^4}\right) \partial^{\mu}\tilde G\sigma^\lambda
\bar{\tilde G} \partial_{\mu}a \partial_{\lambda}a+\left( {2\tilde f\over f^2}\bar {\tilde G}^2(\partial
a)^2+{(\tilde f^*)^2\over 4f^4f_a^2}\tilde G^2\dal\tilde G^2+c.c.\right)~.}}

Let us now analyze the theory~\Lagcompi. The most interesting term is the first operator in the second line
of~\Lagcompi,  $-i
\partial^{\mu}\tilde G\sigma^\lambda \bar{\tilde G} \partial_{\mu}a \partial_{\lambda}a$. In section 3, we have
shown that the coefficient must be non-negative to avoid superluminal propagation. Unitarity provides a slightly
stronger constraint. As in~\nima, the coefficient of the operator $-i \partial^{\mu}\tilde G\sigma^\lambda
\bar{\tilde G}
\partial_{\mu}a \partial_{\lambda}a$ can be related to an integral of a total cross section, therefore once we
include radiative corrections, we expect the coefficient to be strictly positive.  This means that
\eqn\ineqi{{f_a^2\over f^2}-4{|\tilde f|^2\over f^4}>0~,} or equivalently, \eqn\ineqii{2|\tilde f|<f_af~.} Since
$\tilde f$ is just the VEV of the superpotential we conclude with the claimed bound \eqn\ineqiii{2|W|<f_aF~,}
where $F$, as always, is the Goldstino decay constant (which is identical to the total vacuum energy).

As long as~\ineqii\ is satisfied the coefficient of the operator $\bar {\tilde G} ^2\dal \tilde G ^2$ is
positive (and in fact bounded from below by ${1\over 8f^2}$). This is important as one can show that
unitarity demands the coefficient of $\bar {\tilde G} ^2\dal \tilde G ^2$ to be positive.\foot{Similar unitarity
constraints on fermionic vertices were considered in~\AdamsHP.}

A simple consistency check on~\Lagcompi\ is to contrast it with the explicit model in appendix $\aleph$ which at
tree-level satisfies $2|W|=f_aF$. (Of course, quantum corrections will turn it to a strict inequality.) By
plugging into~\Lagcompi\ $2\tilde f=f_af$ we recover precisely the low energy effective action we computed
explicitly in the appendix by integrating out heavy particles.

\newsec{Conclusions}

In this paper we have demonstrated an exact result in theories that break supersymmetry. We have used the
methods of spontaneously broken symmetries along with unitarity bounds. This has led to a bound involving the
VEV of the superpotential, the $R$-axion decay constant and the SUSY breaking scale \eqn\inequalityconc{\vert
\langle W \rangle \vert < {1 \over 2} f_a F~.} This holds in strongly coupled models in which we do not even
know the appropriate variables to describe SUSY breaking macroscopically. A simple (and somewhat surprising at
first sight) corollary of~\inequalityconc\ is that when SUSY is unbroken, even if $R$-symmetry is broken, a
nonzero VEV for the superpotential cannot be generated.

While it is satisfying, in and of itself, that one can prove exact results in theories which spontaneously break
supersymmetry, we would also like to mention a few possible applications and open questions. The VEV of the
superpotential plays a role in supergravity, where it is relevant both for the $R$-axion mass~\BaggerHH\ and the
cosmological constant. Recently, in~\DineSW, it was noted that in theories with {\it discrete} $R$-symmetries,
the requirement of small cosmological constant constrains
 their breaking.  Indeed, in that reference it was noted that discrete $R$-symmetries often lead to approximate, continuous
  symmetries, and it was conjectured that the superpotential is bounded roughly along the lines we have established
 here.  As a result, the potential importance of such symmetries depends on the scale $F$.

In addition, since the bound involves quantities relating $R$-symmetry breaking and SUSY breaking, it is feasible that it may lead to a better understanding of the connection between $R$-symmetry breaking and SUSY breaking~\nelsonseiberg. We have a great deal of evidence that the observations by Nelson and Seiberg are correct even beyond the regime of validity of their analysis. It would be nice to make this more precise.

It would be satisfying to derive the bound we proved in the language of current algebra. This may lead to further exact results in SUSY-breaking theories.
Needless to say, understanding additional general features of supersymmetry-breaking theories is important.

\bigskip
\centerline{\bf Acknowledgements}

We would like to thank N.~Seiberg, D.~Shih and C.~Marcantonini for interesting discussions. We are grateful to T.~Dumitrescu and A.~Katz for helpful comments on the manuscript. The work of MD and GF was
supported in part by the U.S. Department of Energy. The work of ZK was supported in part by NSF grant PHY-0503584. Any opinions, findings, and conclusions or recommendations expressed in this material are those of the author(s) and do not necessarily reflect the views of the funding agencies.

\appendix{$\aleph$}{Low Energy Analysis of a Simple Sigma Model}
This appendix is dedicated to the analysis of the model
\eqn\onefieldapp{K = K(ZZ^*)~,\qquad W
= f Z}
at low energies. We restrict ourselves to the tree approximation. The simple theory~\onefieldapp\ enjoys $R$-symmetry under which $R(Z)=2$. We assume that the K\"ahler potential is such that there exists a SUSY-breaking vacuum at $|z|=v$ where $v$ is nonzero. Therefore, $R$-symmetry is spontaneously broken. The spectrum of the theory thus consists of a massless Goldstino, a massless $R$-axion and a massive real degree of freedom which we, for simplicity, dub the ``Higgs field."

We denote the K\"ahler metric $g_{zz^*}=\partial_Z\partial_{Z^*}K$ and the Christoffel symbols $\Gamma^z_{zz}={g_{zz^*,z}\over g_{zz^*}}$, $\Gamma^{z^*}_{z^* z^*}={g_{zz^*,z^*}\over g_{zz^*}}$. The Lagrangian corresponding to~\onefieldapp\ is given by
\eqn\compapp{\eqalign{& \CL=-g_{zz^*}\partial_\mu Z\partial^\mu Z^*-ig_{zz^*}\bar\psi\bar\sigma^\mu D_\mu\psi+\left({f\over 2}\Gamma^z_{zz}\psi^2+c.c.\right)+{1\over 4}R_{zz^*zz^*}\psi^2\bar\psi^2
-{|f|^2\over g_{zz^*}}~.}}
Here $D_\mu\psi=\partial_\mu \psi+\Gamma^z_{zz}(\partial_\mu Z)\psi$. Since the K\"ahler potential is only a function of $ZZ^*$, so is the metric $g_{zz^*}$. By assumption there is a vacuum at $|z|=v$ so we substitute $z=(v+h)e^{2ia}$ and we see that $g_{zz^*}$ is independent of $a$. In addition, the existence of a vacuum for $h=0$ implies that in an expansion around the vacuum
\eqn\metricapp{g_{zz^*}=\alpha-{\beta\over v^2} h^2+\cdots~,}
where $\alpha$, $\beta$ are dimensionless numbers that can be easily determined given a specific model. From this we can also read out the expansion of the Christoffel symbols and curvature around the vacuum
\eqn\Chrisapp{\Gamma^z_{zz}=-{\beta\over \alpha v^2}he^{-2ia}+\cdots~,\qquad \Gamma^{z^*}_{z^*z^*}=-{\beta\over \alpha v^2}he^{2ia}+\cdots~,\qquad R_{zz^*zz^*}=-{\beta\over 2v^2}+\cdots~.}

We recall that the Higgs field $h$ is massive, and in order to understand the couplings in the IR we should integrate it out. The first approximation is to set it to zero, but we want to read out the low energy effective action more carefully, so we solve the equations of motion of the theory~\compapp\ at the leading nontrivial order in the number of Goldstinos, axions and derivatives. The full Lagrangian~\compapp\ truncated to contain only the leading order terms in the Higgs field, according to~\metricapp,\Chrisapp\ is
\eqn\truncatedLag{\eqalign{&-\CL={|f|^2\over\alpha}+\alpha(\partial h)^2+4\alpha v^2(\partial a)^2+i\alpha\bar\psi\bar\sigma^\mu\partial_\mu\psi+8\alpha v h(\partial a)^2+{2\beta\over v}h\partial_\mu a\bar\psi\bar\sigma^\mu\psi\cr&+\left({f\beta\over 2\alpha v^2}he^{-2ia}\psi^2+c.c.\right)+{\beta\over 8v^2}\psi^2\bar\psi^2+{|f|^2\beta\over \alpha^2 v^2}h^2~.}}
The resulting equation of motion for the heavy Higgs field is solved by
\eqn\soleomapp{h=-\left({\alpha f\over 4|f|^2}e^{-2ia}\psi^2+c.c.\right)-{\alpha^2v\over |f|^2}\partial_\mu a\bar\psi\bar\sigma^\mu\psi-{4\alpha^3 v^3\over \beta|f|^2}(\partial a)^2~.}
We can now plug it back to the action to read out the effective Lagrangian for the axion and Goldstino. We only keep the leading terms describing self interactions of the axion and Goldstino as well as the leading operator connecting the axion with the Goldstino. The final result is
\eqn\finaleff{\eqalign{&\CL_{eff}=-{|f|^2\over\alpha}-4\alpha v^2(\partial a)^2-i\alpha\bar\psi\bar\sigma^\mu\partial_\mu\psi+\alpha \left({\alpha \over 4f^*}\psi^2+c.c.\right)\partial^2 \left({\alpha \over 4f^*}\psi^2+c.c.\right)\cr&
+{16\alpha^4v^4\over \beta|f|^2}(\partial a)^4+\left({2v\alpha^2\over f^*}\psi^2(\partial a)^2+c.c.\right)~.}}
In addition to the expected kinetic terms and vacuum energy, we wrote down the leading interactions of the Goldstino with itself, the axion with itself and the leading operator that connects them. Note that the self interaction of the axion has a positive sign, which is guaranteed by unitarity~\nima. A similar comment holds for the term $\psi^2\partial^2(\bar\psi^2)$ which appears in the self interaction of the Goldstino.

The Lagrangian~\finaleff\ takes a more natural form once the Goldstino is canonically normalized (in spite of the change of variables we retain the notation) and we express everything in terms of the decay constant $f_a^2=4\alpha v^2$ and SUSY breaking scale $F=|f|/\sqrt \alpha$. Without loss of generality, we also assume that $f$ is real. We get
\eqn\finaleffi{\eqalign{&\CL_{eff}=-F^2-f_a^2(\partial a)^2-i\bar\psi\bar\sigma^\mu\partial_\mu\psi+ \left({1 \over 4F}\psi^2+c.c.\right)\partial^2 \left({1 \over 4F}\psi^2+c.c.\right)\cr&
+{\alpha f_a^4\over \beta F^2}(\partial a)^4+\left({f_a \over F}\psi^2(\partial a)^2+c.c.\right)~.}}
We see that two of the interaction terms depend only on $F,f_a$ which suggests that they are associated to universal terms in the low energy effective action. On the other hand, the axion quartic interaction
depends on the details of the high energy physics (e.g. the parameter $\beta$). We show in section~4 that these facts follow from studying nonlinear realizations of broken $R$-symmetry and supersymmetry.

\listrefs

 \end